\documentclass[12pt]{article}
\pdfoutput=1
\usepackage{amsmath,amssymb}

\usepackage{graphicx}
\usepackage{color}
\usepackage[bookmarksnumbered=true,colorlinks=true,linkcolor=black,citecolor=black]{hyperref}
\newlength{\extraspace}
\newlength{\extraspaces}
\def\bsklength{.8mm} 

\newcommand{\beq}{\begin{equation}}
\newcommand{\eeq}{\end{equation}}

\newcommand{\bseq}{\addtocounter{subeqno}{1}\begin{subequations}}
\newcommand{\eseq}{\end{subequations}}

\newcommand{\BZ}{{\mathbb Z}}

\font\mathscript=eusm10 at 12pt
\font\mathscripts=eusm7
\font\mathscriptss=eusm5
\newfam\mathscri
\textfont\mathscri=\mathscript
\scriptfont\mathscri=\mathscripts
\scriptscriptfont\mathscri=\mathscriptss
\def\mathscr#1{{\fam\mathscri\relax#1}}

\font\mathfrakt=eufm10 at 12pt
\font\mathfrakts=eufm7
\font\mathfraktss=eufm5
\newfam\mathfraki
\textfont\mathfraki=\mathfrakt
\scriptfont\mathfraki=\mathfrakts
\scriptscriptfont\mathfraki=\mathfraktss
\def\mathfrak#1{{\fam\mathfraki\relax#1}}

\def\CO{{\cal O}}

\renewcommand{\tilde}{\widetilde}

\renewcommand{\bar}{\overline}

\newcommand{\e}{{\rm e}}
\newcommand{\tr}{{\rm tr}}
\newcommand{\Tr}{{\rm Tr}}

\newcommand{\gsim}{\gtrsim}

\begin{document}
\setcounter{page}{0}
\addtolength{\baselineskip}{\bsklength}
\thispagestyle{empty}
\renewcommand{\thefootnote}{\fnsymbol{footnote}}        

\begin{flushright}
arXiv:1110.6919 [hep-ph]\\
\end{flushright}
\vspace{.4cm}

\begin{center}
{\Large
{\bf{Techni-Chiral-Color}}}\\[1.2cm]
{\rm Thomas W. Kephart\footnote{tom.kephart@gmail.com} 
and 
HoSeong La\footnote{hsla.avt@gmail.com} 
}
\\[3mm]
{\it Department of Physics and Astronomy,\\[1mm]
Vanderbilt University,,\\[1mm]              
Nashville, TN 37235, USA} \\[1.5cm]

\vfill
{\parbox{15cm}{
\addtolength{\baselineskip}{\bsklength}
\noindent
Chiral Color is extended by incorporating Technicolor,
which induces dynamical breaking of the Electroweak symmetry 
as well as Chiral Color to Quantum Chromodynamics.
Gauge anomalies are cancelled by introducing two generations of 
technifermions, and the fourth generation of quarks and leptons is required.
Each technifermion generation is coupled to only two Standard Model generations
by the Yukawa interaction. 
Various phenomenological implications are explained.

}
}


\end{center}
\noindent
\vfill


\newpage
\setcounter{page}{1}
\setcounter{section}{0}
\setcounter{equation}{0}
\setcounter{footnote}{0}
\renewcommand{\thefootnote}{\arabic{footnote}}  
\newcounter{subeqno}
\setcounter{subeqno}{0}
\setlength{\parskip}{2mm}
\addtolength{\baselineskip}{\bsklength}

\pagenumbering{arabic}



The model introduced here is a hybrid between Chiral Color (CC)\cite{fg,PS} 
and Technicolor (TC)\cite{tc,etc}.
Neither of these have any evidence of their existence. 
However, there are good reasons why these could be the immediate future
of new physics beyond the Standard Model (SM).

First, the nature of chirality has been fascinating us since the discovery of
parity violation and the V-A theory. The correct identification of quarks and
leptons in the SM based on anomaly cancellation proves
the value of the chiral nature of the Electroweak theory. Yet, eventually
the low energy world of unbroken symmetry is vector-like. 
One cannot help but raising the further question of why Quantum Chromodynamics 
(QCD) (i.e. SU(3)$_{\rm C}$)
is vector-like, while part of the Electroweak theory (i.e. SU(2)$_L$) is 
chiral. In fact, as a global flavor symmetry, chiral symmetry is introduced in
QCD to explain the origin of (light) quark masses.
We can go one step further and ask if QCD itself is a result of 
spontaneous breaking of local chiral symmetry. 
Indeed this question was asked
before and the model constructed is known as the Chiral Color\cite{fg}.

Second, the existence of the Higgs to provide the electroweak symmetry 
breaking (EWSB) in terms of an elementary scalar is still elusive 
and it is possible that we may face the situation of no discovery. 
If there is no Higgs, the most obvious alternative is
clearly dynamical symmetry breaking\cite{dysb} 
and there are active investigations going on
in the context of TC\cite{fs,ken,Hill:2002ap}.

Furthermore, we will break both CC and EW symmetry dynamically
at the same time. Then we can resolve many outstanding issues.
To name a few, formulation of a realistic CC with sufficiently heavy
axigluons, heavy top-quark mass and fermion mass hierarchy,
small mixing between top-quark and others in the 
Cabibbo-Kobayashi-Maskawa (CKM) mixing matrix, suppression of 
of the flavor changing neutral current (FCNC), etc.
The cost of doing this is the addition of two technifermion generations
and a fourth generation of quarks and leptons.


The model we consider is based on the (relevant) gauge group\footnote{The 
gauge group is actually $G\times{\mbox SU(2)}'$.
However, no matter fields carry SU(2)$'$ charges so that SU(2)$'$ is
decoupled from the matter sector. 
The role of SU(2)$'$ is to be clarified later.}
\beq
G=
{\rm SU}(2)_{\rm TC} \times {\rm SU}(3)_L \times {\rm SU}(3)_R
\times {\rm SU}(2)_L \times {\rm U}(1)_Y,
\eeq
where quarks and leptons are ${\rm SU}(2)_{\rm TC}$ singlets 
and given by
\begin{align}
Q: &(1,3,1,2, 1/6)_L,(1,1,3,1,2/3)_R,(1,1,3,1,-1/3)_R, \nonumber\\
L: &(1,1,1,2,-1/2)_L,(1,1,1,1,-1)_R
\end{align}
where we suppress family indices.

In the CC case there is an anomaly due to the chiral nature of SU(3)'s.
There are various ways of canceling the SU(3) anomaly by introducing
additional fermions, as  presented in \cite{fg}. However,
there is also a rather trivial way of canceling
SU(3) anomalies by simply adding anti-generations.
One may think this is not quite an innovative idea, but if we
treat these anti-generations as technifermions with
additional strong ${\rm SU}(2)_{\rm TC}$ interaction,
then it gets interesting. (See \cite{Kuo:1987jx} for another way of
incorporating TC with CC.) 
So, we are led to include
\begin{align}
{T_Q}^*&: (2,\bar{3},1, 2, -1/6)_L,(2,1,\bar{3},1,-2/3)_R,
(2,1,\bar{3}, 1, 1/3)_R,\nonumber\\
T_L&: (2,1,1,2,1/2)_L, (2,1,1,1,0)_R, (2,1,1,1,1)_R,
\end{align}
where the asterisk is added to indicate $\bar{3}$.
Notice that we still define the electric charge 
as $Q_{\rm EM}=I_{L3}+Y/2$ because 
${\rm SU}(2)_{\rm TC}$ will remain unbroken but confined.
Since the $T_Q$'s are doublet under ${\rm SU}(2)_{\rm TC}$, 
to cancel SU(3) anomalies we need to match one generation of technifermions 
with two generations of $Q, L$'s. 
Therefore, in total we have two generations of technifermions and
four generations of $Q, L$, hence the fourth generation is needed.
This distinguishes the top-quark (and $t'$) generation
from the two light generations as in the Topcolor 
model\cite{Hill:1991at,topcolor}.


The $T_Q$'s and $T_L$'s will condense due to strong 
${\rm SU}(2)_{\rm TC}$ interactions,
and this should break the EW symmetry as well as CC.
We will count Nambu-Goldstone bosons (NGB) slightly differently,
but it is equivalent to a total of 255 NGB's due to the
$SU(16)_L\times SU(16)_R$ chiral symmetry breaking down to $SU(16)_V$
\footnote{Note that the global symmetry is not SU(32) due to CC charges.
The extra 240 condensates which could have appeared under
enhanced global symmetry SU(32) breaking to Sp(32) are not Lorentz scalars, 
although they are SU(2) singlets. 
So they should not be counted as NGB's in our case.}.
Because of four different colored $T_Q$'s, 
there are sixteen condensates formed by $T_Q$'s under SU(2)$_{\rm TC}$. 
However, to be more precise,
all of these have $(3,\bar{3})$ degrees of freedom according to CC.
Upon twisting CC to QCD times the axial SU(3), these degrees of freedom
separate into SU(3) singlets and octets, 144 in total. 
In addition to these, there are another fifteen condensates formed by
$T_L$'s under SU(2)$_{\rm TC}$ and 
sixteen colored condensates of $\langle\bar{T_Q}T_L\rangle$ 
and their conjugates, in total 111. 
Out of sixteen SU(3) octets, one will be eaten to break 
SU(3)$_L\times$SU(3)$_R$ to the SU(3)$_{\rm C}$ of QCD to make 
eight axigluons massive. 
The remaining fifteen octets will form condensates due to SU(3)$_{\rm C}$
which could have properties similar to glueballs in QCD after confinement.
Three combinations of $\langle\bar{T_L}T_L\rangle$ and the singlet 
$\langle\bar{T_Q}T_Q\rangle$ will be eaten to break 
the electroweak symmetry. The remainders are 
psuedo-Nambu-Goldstone bosons (PNGB) and techniaxions\cite{Hill:2002ap}.

To match the EWSB scale the technipion (i.e. singlet) decay constant
is given by 
$
F_{\pi_{\rm TC}}=(246\ {\rm GeV})/\sqrt{8}= 87\ {\rm GeV},
$
which leads to the correct weak boson masses\cite{ken,Hill:2002ap}.
Notice that, for axigluons, the octet decay constant $F_{g_{\rm TC}}$
does not have to be the same as $F_{\pi_{\rm TC}}$, depending on 
the detail of flavor symmetry breaking. 
They may be related, but we do not have any experimental data to use 
to fix parameters needed to specify the relationship at this moment. 
Since we can construct an effective lagrangian in which
(color) singlet and octet technimesons have independent kinetic energy 
terms with their own decay constants, e.g. 
$(F_{\pi_{\rm TC}}^2/4)\tr|D_\mu U_1|^2
+(F_{g_{\rm TC}}^2/4)\Tr|D_\mu U_8|^2$, where $U_1$ is given in terms of
isospin triplet $\pi_T$ as 
$U_1\equiv {\rm exp}(i\sigma\cdot \pi_T/F_{\pi_{\rm TC}})$ and 
$U_8$ is in terms of color octet $\Pi_8$ as
$U_8\equiv {\rm exp}(i\lambda\cdot \Pi_8/F_{g_{\rm TC}})$,
we can safely assume they are independent at this moment.
So, the axigluon mass is given by
\beq
m_{g_A} 
\sim \frac{1}{2}g_s\sqrt{8C_2}F_{g_{\rm TC}}
\eeq
where $g_s$ is the QCD coupling constant
and $C_2=3$ is the second Casimir invariant
for the adjoint representation of SU(3).
If we choose $F_{g_{\rm TC}}\sim\Lambda_{\rm TC}/2\sim 0.5$ TeV, 
then the axigluon mass becomes about 1.2 TeV.


We will generate fermion masses without introducing Extended Technicolor 
(ETC)\cite{etc}, but in terms of strongly interacting scalars 
and sterile scalars. 
(Generating mass via color singlet scalars in TC
was investigated in \cite{stc,georgi,kagan1}.)

The necessary scalar field for generating quark masses is
\beq
\Phi: (2,1,1,1,0).
\eeq
which only carries TC charge.
If this scalar field interacts with fermions according to Yukawa 
couplings of 
\beq
\lambda_{Q}\Phi_A\bar{Q}^a {T_Q}^A_a + {\rm h.c.},
\eeq
where the label `$A$' is an ${\rm SU}(2)_{\rm TC}$ index 
and `$a$' is a CC index,
then condensations of $T_Q$'s and $\Phi$'s will generate masses. 
Motivated by the anomaly cancellation structure, we assume that 
the first (second) generation of $T_Q$ couples only
to the first and second (third and fourth) generations of $Q$.
This assumption can be justified either
by imposing restrictions on Yukawa coupling constants
or by imposing a discrete symmetry. 
In the case of the discrete symmetry $\Gamma$, 
we assume that Yukawa couplings respect $\Gamma$, 
while the TC gauge interaction is allowed to violate $\Gamma$
because $T_{Q_1}$ and $T_{Q_2}$ are identified as weak eigenstates,
so that there is no reason why TC should have the same eigenstates.
The simplest example is $\Gamma=\BZ_2$ such that different technifermions
have different $\BZ_2$-parities and others are assigned accordingly, then
no Yukawa coupling mixing will be allowed.
The scalar $\Phi$ is also assumed to be non-self-interacting for simplicity 
and it will be confined.

The mass matrix for, say, up-quarks, is a $4\times 4$ matrix with $2\times 2$
blocks that can be expressed as
\beq
M_Q =(M_{IJ}),
\eeq
where $I,J$ identify the technifermion generations and each block is
given by $2\times 2$ matrix of the form
\beq
M_{IJ}=(\sum_{\ell=1}m^{(\ell)}_{ij}),
\eeq
where $i,j$ identify the quark generations. Notice that for $I(J)=1$
$i(j)=1,2$, while for $I(J)=2$, $i(j)=3,4$.

\begin{figure}[ht] 
   \centering
   \vskip-4pt
   \includegraphics[width=2.5in]{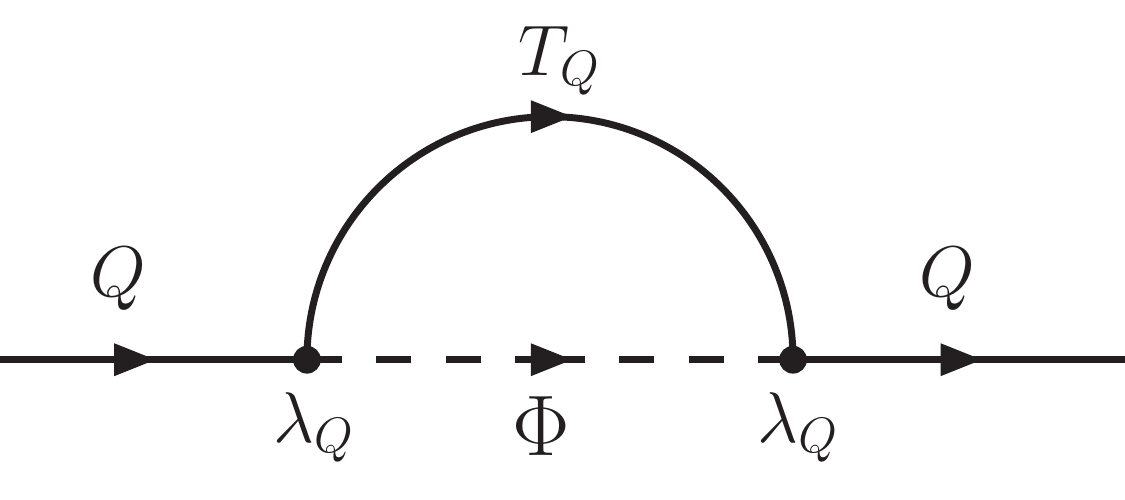} 
   \parbox{0.8\textwidth}{\vskip-6pt
   \caption{Feynman diagram for diagonal block quark mass at the lowest order.}
   \label{fig:f1}}
\end{figure}

\begin{figure}[ht] 
   \centering
   \vskip-12pt
   \includegraphics[width=2.5in]{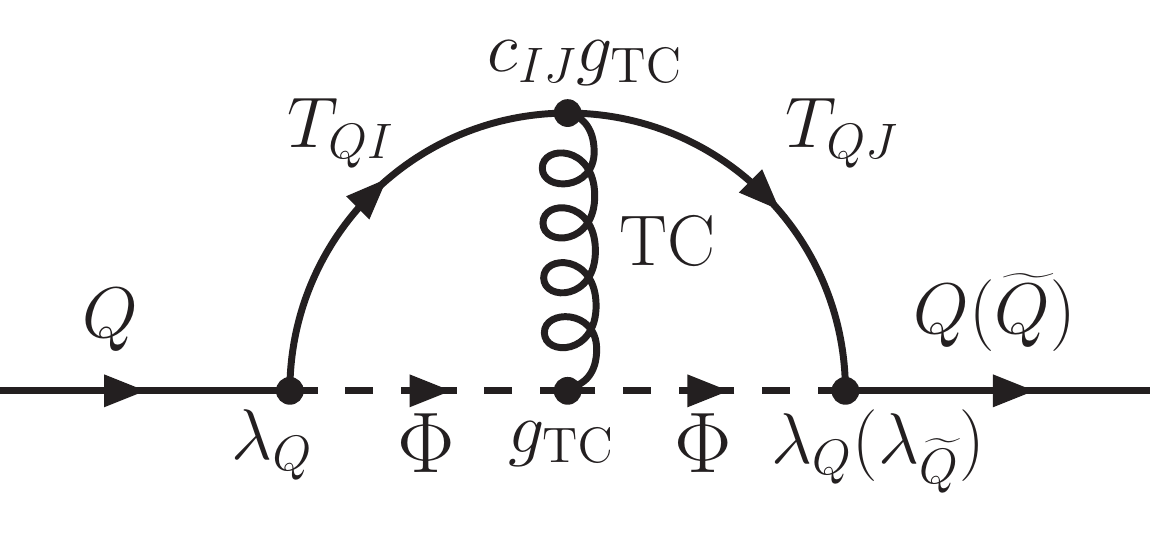} 
   \parbox{0.8\textwidth}{\vskip-6pt
   \caption{An example of Feynman diagram for higher order contributions.
   Those inside brackets are for off-diagonal-block quark mass. 
   $c_{IJ}$ indicate technifermion mixingings.}
   \label{fig:f3}}
\end{figure}

We will consider only the two leading order
contribution to the diagonal block
\beq
M_{II}=(m^{(1)}_{ij}+m^{(2)}_{ij}).
\eeq
The condensations of $T_Q$'s and $\Phi$'s in the dimension eight operator 
of an effective action represented by Fig.\ref{fig:f1} 
lead to the first term
\beq
\label{e:e11}
m^{(1)}_{ij}
\sim \lambda_i\lambda_j
{\langle\Phi^2\rangle \langle T_{QI}^2\rangle\over \Lambda_{\rm TC}^4}
\sim {1\over 16\pi^4}\lambda_i\lambda_j m_{T_{QI}}.
\eeq
A contribution for the diagonal block 
at the two-loop level generated by Fig.\ref{fig:f3} gives
\beq
\label{e:e12}
m^{(2)}_{ij}\sim {1\over 256\pi^8}
c_{II}g_{\rm TC}^2\lambda_i\lambda_j 
{m^2_{T_{QI}}\over\Lambda_{\rm TC}},
\eeq
where $c_{II}=1$.

Similarly, the leading order of the off-diagonal-block components is generated 
at the two-loop level (Fig.\ref{fig:f3} with those fields inside the bracket) 
from a dimension sixteen operator such that
\beq
\label{e:e13}
M_{IJ}=(m_{ij}^{(2)})\sim \left({1\over 256\pi^8}
c_{IJ}g_{\rm TC}^2\lambda_i\lambda_j 
{m_{T_{Q1}} m_{T_{Q2}}\over\Lambda_{\rm TC}}\right),
\eeq
where $c_{\rm IJ}$ is the technifermion mixing, 
hence suppressed by order $m_{T_Q}/\Lambda_{\rm TC}$ compared 
to the diagonal ones.
For $c_{IJ}\neq 0$ when $I\neq J$, 
we need $T_{Q1}$ and $T_{Q2}$ mixing for their couplings to TC gauge fields. 
This is because $T_{Q1}$ and $T_{Q2}$ are weak eigenstates, 
hence there is no reason for them
to interact with SU(2)$_{\rm TC}$ in the same way.
This mixing indicates that SU(2)$_{\rm TC}$ is not the entire gauge symmetry, 
but in fact part of a larger gauge symmetry as follows: 
Let SU(2)$_{\rm TC}$ be a (properly) twisted 
part as in $\mbox{SU(2)}_1\times\mbox{SU(2)}_2
\simeq\mbox{SU(2)}_{\rm TC}\times\mbox{SU(2)}'$
with coupling constants given by
$g_1=(1+c_{12})g_{\rm TC}$ and $g_2=(1-c_{12})g_{\rm TC}$, where SU(2)$'$
is confining and no matter fields carry SU(2)$'$ charges 
so that SU(2)$'$ decouples, then we only have SU(2)$_{\rm TC}$ 
coupled to matter in the $\Lambda_{\rm TC}$ region. 
So, the gauge invariance under $\mbox{SU(2)}_1\times\mbox{SU(2)}_2$
can be demonstrated by untwisting as follows.
Let $T_Q=(T_{Q1}+T_{Q2})/\sqrt{2}$ and $\tilde{T}_Q=(T_{Q1}-T_{Q2})/\sqrt{2}$ 
such that $g_1\bar{T_Q}A^{(1)}_\mu T_Q 
+ g_2 \bar{\tilde{T}_Q}A^{(2)}_\mu\tilde{T}_Q$,
then this leads to gauge invariance. Rewrite in terms of $T_{Q1}$ and $T_{Q2}$
and, since they only carry SU(2)$_{\rm TC}$ charges, identifying
$A^{\rm TC}_\mu=A^{(1)}_\mu=A^{(2)}_\mu$, we get the technifermion mixing terms.
Notice that $T_Q(\tilde{T}_Q)$ couples to SU(2)$_1$($\mbox{SU(2)}_2$) only.
This mixing is crucial to generate the desired mass matrix and 
CKM-like mixing, and in fact is the origin 
of the mixing in the SM in this context.
In the gauge sector the gauge fields of SU(2)$'$ behave like vector 
``matter'' fields with respect to SU(2)$_{\rm TC}$ upon twisting 
$\mbox{SU(2)}_1\times\mbox{SU(2)}_2$\cite{La:2003jm}. So,
the evidence of SU(2)$'$ will show up in the TC gauge sector only,
e.g., to modify Ward identities of SU(2)$_{\rm TC}$ in some cases.

\begin{figure}[t] 
   \centering
   \hskip.4in
   \includegraphics[width=1.2in]{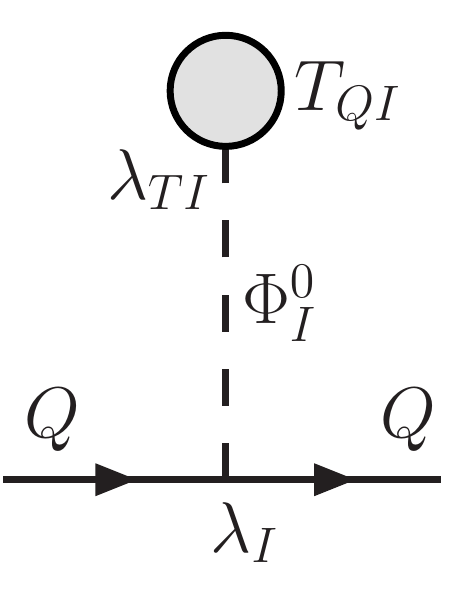} 
   \parbox{0.8\textwidth}{\vskip-6pt
   \caption{Tadpole diagram for flavor diagonal masses.}
   \label{fig:f5}}
\end{figure}

The mass matrix given in terms of Eqs.(\ref{e:e11})-(\ref{e:e13})
has two zero eigenvalues.
To avoid these zero eigenvalues we need to add a flavor diagonal term
and the tadpole contribution can take care of this. 
So, we are led to introduce two sterile scalar fields in the same spirit
of grouping generations as prescribed (one is sufficient too
if we allow much larger hierarchy of Yukawa coupling constants)
\beq
\label{e:s0}
\Phi^0_I : (1,1,1,1,0)
\eeq
such that they only interact with (techni)fermions via Yukawa couplings as
\beq
\lambda_{TI}\Phi^0_I\bar{T_{QI}} T_{QI},
\quad \lambda_i\Phi^0_I\bar{Q}_i Q_i.
\eeq
Notice that $\Phi^0_1$ interacts only with $Q_{1,2}$, etc., as before
due to restrictions on Yukawa coupling constants.
Then the tadpole contribution from Fig.\ref{fig:f5} generates
the diagonal contribution
\beq
m_{ii}\sim {1\over 2\pi^2}\lambda_i\lambda_{TI}m_{T_{QI}}
\left({\Lambda_{\rm TC}\over m_{\Phi^0_I}^2}\right)^2,
\eeq
where their masses are free parameters in this context.
However, the mass can be easily generated dynamically by assuming
another strong interaction, e.g. SU(2)$'$ mentioned before, 
whose charge only these scalars carry and the radial components become
$\Phi^0_I$.

The down-quarks mass matrix can be generated similarly and the difference
compared with the up-quarks mass matrix
will be due to the difference in Yukawa couplings and technifermion masses.


Let $V_u$ and $U_d$ be the matrices which diagonalize the up-quarks 
and down-quarks mass matrices, respectively. Since the initial
mass matrices are of block-form, 
$V_u$ and $U_d$ will also naturally be of block-form. However
different blocks can have different phase ambiguities so that
mixing matrix can have different phases for different blocks.
Hence, we can construct the mixing matrix as 
\beq
K=V_u^\dagger U_d,
\eeq
where
\beq
V_u =
\begin{pmatrix}
\e^{i\delta_{11}}V_{11} & \e^{i\delta_{12}}V_{12} \\
\e^{-i\delta_{12}}V_{21} & \e^{i\delta_{22}}V_{22} \\
\end{pmatrix},\
U_d =
\begin{pmatrix}
\e^{i\tilde\delta_{11}}U_{11} &\e^{i\tilde\delta_{12}}U_{12} \\
\e^{-i\tilde\delta_{12}}U_{21} &\e^{i\tilde\delta_{22}}U_{22} \\
\end{pmatrix}.
\eeq
In principle, a $4\times 4$ unitary matrix can have three independent phases.
So, we identify $\delta_{11}=\tilde\delta_{11}$ and
$\delta_{22}=\tilde\delta_{22}$, then there remain three independent
phases, namely, $\delta_{12}-\tilde\delta_{12},
\delta_{11}-\delta_{12},$ and $\delta_{22}+\delta_{12}$.
The first $3\times 3$ submatrix of this $4\times 4$ matrix can be 
identified as the usual CKM matrix, albeit parametrized in terms of the
usual Eulerian angles. Needless to say, there is a unitary transformation 
changing this to the standard CKM parametrization.
Due to the structure of the mass matrix,
the off-diagonal-blocks are smaller by order of
$c_{12}m_{T_Q}/\Lambda_{\rm TC}$ than the diagonal blocks,
and so is the mixing matrix. 
This can easily explains why in the CKM matrix, the
top mixings to the first two generations are smaller by order of $10^{-2}$
than the mixing of the first with the second generations.
This in turn explains the suppression of FCNC.


To leading order the diagonalized masses are 
(assuming all Yukawa couplings are of order unity) of the form
\begin{align}
&m_u:m_c: m_t: m_{t'} \nonumber\\
\sim\ & m_{T_{Q1}}
\left({\Lambda_{\rm TC}\over m_{\Phi^0_1}}\right)^2
: m_{T_{Q1}}
:m_{T_{Q2}}\left({\Lambda_{\rm TC}\over m_{\Phi^0_2}}\right)^2
: m_{T_{Q2}}.
\end{align}
There are seven unknowns including $m_{t'}$
and only four relations, so there are three
free parameters (in addition to Yukawa coupling constants).
Using known masses of quarks and assuming $\Lambda_{\rm TC}\sim 1\ {\rm TeV}$,
we can immediately estimate $m_{\Phi_1^0}\sim 23$ TeV for $m_u\sim 2.5$ MeV.
If we further assume $m_{T_{Q2}}\sim m_{t'}\gsim 300$ GeV, 
then $m_{T_{Q1}}\sim m_c$
and $m_{\Phi_2^0}\sim 1.3$ TeV. This in turn will fix the down-quarks
masses with suitable Yukawa coupling constants.
The mass of the strongly interacting scalar $\Phi$ is missing
here because the corresponding contribution only shows up at higher orders.
These are not strong constraints, and there is much room for variations,
including different Yukawa coupling constants.
For example, if we have only one sterile scalar coupling universally
to all (techni)fermions, then $\lambda_t^0/\lambda_u^0\sim m_t/m_u$ as
in SM, etc., can generate the same mass hierarchy. 
But we find this quite uninspiring.
On the contrary, in our model it is easy to generate the heavy top quark mass
and explain the known mass hierarchies in terms of Yukawa coupling constants
of similar order of magnitude.


For lepton masses, we can use the same strongly 
interacting $\Phi$ such that the necessary Yukawa couplings are given by
\beq
\lambda_L\Phi_A\bar{L} T_L^{*A} +{\rm h.c},
\quad 
\eeq
where the label `$A$' is an ${\rm SU}(2)_{\rm TC}$ index.
The flavor diagonal contribution is again 
due to the sterile scalars given in Eq.(\ref{e:s0}),
then, the lepton masses can be generated accordingly.
The difference compared to the quark masses 
is now that the lepton masses are generated by
condensation of technileptons $T_L$'s so that it can easily
accommodate the quark-lepton mass hierarchy.
For given $\Lambda_{\rm TC}$ and $m_{\Phi^0_I}$ estimated from the
quark mass hierarchy, we can generate the lepton mass hierarchy
based on the limit on the fourth generation leptons,
without unreasonably small or large Yukawa coupling constants.
Lepton mixing matrix can be similarly constructed from this lepton
mass matrix. 

In fact, the structure we have introduced in this letter is fairly generic.
As long as one technifermion generation (with or without technileptons)
couples to two SM generations as prescribed, all phenomenological
outcomes are similar. We can do this even without CC, although
less motivated coupling one technifermion generations to two
SM model generations. We suspect there might be a larger framework
from which this structure can be inherited and justified.

The electroweak precision constraints ruled out QCD-like TC models
with degenerate technifermion doublets, but
we have good reasons why our model should be safe.  
The details will be presented elsewhere\cite{mine}.
First, it is known that massive scalars and extra nondegenerate
heavy fermions can contribute to meet the precision 
data\cite{georgi,Peskin:1991sw,Dobrescu:1997kt,Dobrescu:1998ci,pdg}.
We have three massive scalars and fourth SM generation, in addition to
technifermions that can be non-degenerate.
Second, the precision data test is based on TC$\times$SM,
not TC$\times$CC$\times$EW. So,
it is possible that CC may modify the outcome.
Third, strictly speaking, our model is not QCD-like because of
technifermion mixing.

Since $T_{Q1}$ or $T_{L1}$ is the lightest technifermion, it is a good place 
to look for a signal to distinguish this model.
Both interact with known world particles above the TC scale, 
while emitting a strongly interacting scalar, which could lead to a monojet.
One may think their masses may be too low, but it is acceptable because 
neither of them will show up as a quark or a lepton
below $\Lambda_{\rm TC}$ due to SU(2)$_{\rm TC}$.
What we observe at low energy will be their (doubly for $T_{Q1}$)
confined objects, whose mass can be quite high beyond the present measured 
scale.
Furthermore, the mass can always be raised by increasing $\Lambda_{\rm TC}$
or adjusting Yukawa coupling constants.

Both strongly interacting and sterile neutral scalar fields 
could be candidates for the dark matter. 
Their masses can be at least of order TeV, which is well beyond the current
limit of about 500 - 600 GeV set by LHC\cite{higgscern}. 
These scalars interact with the known world particles only in terms of Yukawa 
interactions. 
The lowest level flavor changing effective Yukawa couplings generated 
by technifermion condensation appear only at two-loop level, and their
contribution to FCNC amplitudes below $\Lambda_{\rm TC}$
is suppressed at least, even for tree-level Yukawa couplings of order unity, by 
$\CO((m_{T_Q}/2\pi\Lambda_{\rm TC})^{2(\ell+1)})$, where $\ell=2$ is 
the lowest number of TC loops needed.
Hence, the mass bounds on these scalars are even lower 
by $\CO((m_{T_Q}/2\pi\Lambda_{\rm TC})^3)$ 
than those of Yukawa coupled non-Higgs scalars'.
The highest mass bound of the latter from flavor physics is about 
1 TeV\cite{sbound}.
Then the bound on scalars in our case is at most about one tenth of that,
which is lower than the bound set by LHC.
Since we have SU(2)$_{\rm TC}$,
we do not expect the technibaryon problem\cite{Chivukula:1989qb}.

The mixing matrix we proposed here accommodates the CKM matrix, 
yet goes further so that it is a good place to look for
clues of physics beyond the SM. There are two additional CP-violating
phases involving the fourth generation, hence the model allows
much more room for CP-violation.
The SM has a difficulty explaining the baryon asymmetry
due to insufficient CP-violation. Our current work certainly opens the door
to resolving this issue.

One shortcoming of our model is that we still cannot avoid the large number
of PNGB's and techniaxions as in most of TC models. The only way to reducing
the number is reducing that of technifermions. The possibility of eliminating
technileptons based on \cite{La:2003jm} is under investigation. 
Also, it will be nice to generate flavor diagonal masses without 
sterile scalar fields so that we can reduce free parameters.

We have presented our basic ideas briefly in this letter,
but more details and variant models will be presented elsewhere\cite{mine}.

HL thanks the theory group of the Department of Physics \& Astronomy
at Vanderbilt University for their hospitality while this work is
completed. We thank Alex Kagan for bringing references
\cite{kagan1}, \cite{Dobrescu:1997kt} and \cite{Dobrescu:1998ci} 
to our attention.
TWK is supported by US DOE grant number DE-FG05-85ER40226.

\end{document}